\documentclass[twocolumn,secnumarabic,amsmath,amssymb,nobibnotes, aps, prl,superscriptaddress,]{revtex4-2}
\usepackage{graphicx}
\usepackage{dcolumn}
\usepackage{siunitx}
\usepackage{multirow}
\usepackage{color}
\usepackage{threeparttable}

\usepackage[mathlines]{lineno}

\begin{document}

\title{New Ground State in ${}^{149}$La Removes \\ Two-Neutron-Separation-Energy Anomaly in Lanthanum Isotopes}

\author{S.~Kimura}
\email[]{sota.kimura@kek.jp}
\affiliation{Wako Nuclear Science Center, Institute of Particle and Nuclear Studies, High Energy Accelerator Research Organization, Wako 351-0198, Japan}

\author{M.~Wada}
\affiliation{Wako Nuclear Science Center, Institute of Particle and Nuclear Studies, High Energy Accelerator Research Organization, Wako 351-0198, Japan}
\affiliation{Advanced Energy Science and Technology GuangDong Laboratory, Huizhou 516007, China}
\affiliation{Institute of Modern Physics, Huizhou branch, Chinese Academy of Science, Huizhou 516000, China}

\author{H.~Haba}
\affiliation{RIKEN Nishina Center for Accelerator-Based Science, Wako 351-0198, Japan}

\author{Y.~Hirayama}
\affiliation{Wako Nuclear Science Center, Institute of Particle and Nuclear Studies, High Energy Accelerator Research Organization, Wako 351-0198, Japan}

\author{H.~Ishiyama}
\affiliation{RIKEN Nishina Center for Accelerator-Based Science, Wako 351-0198, Japan}

\author{Y.~Ito}
\affiliation{Wako Nuclear Science Center, Institute of Particle and Nuclear Studies, High Energy Accelerator Research Organization, Wako 351-0198, Japan}

\author{T.~Niwase}
\altaffiliation[Present address: ]{Department of Physics, School of Science and Technology, Meiji University, Kawasaki, 214-8571, Japan
}
\affiliation{Department of Physics, Kyushu University, Fukuoka 819-0395, Japan}

\author{M.~Rosenbusch}
\affiliation{RIKEN Nishina Center for Accelerator-Based Science, Wako 351-0198, Japan}

\author{P.~Schury}
\affiliation{Wako Nuclear Science Center, Institute of Particle and Nuclear Studies, High Energy Accelerator Research Organization, Wako 351-0198, Japan}

\author{H.~Ueno}
\affiliation{RIKEN Nishina Center for Accelerator-Based Science, Wako 351-0198, Japan}

\author{Y.X.~Watanabe}
\affiliation{Wako Nuclear Science Center, Institute of Particle and Nuclear Studies, High Energy Accelerator Research Organization, Wako 351-0198, Japan}

\author{Y.~Yamanouchi}
\affiliation{Department of Physics, Kyushu University, Fukuoka 819-0395, Japan}
\affiliation{RIKEN Nishina Center for Accelerator-Based Science, Wako 351-0198, Japan}
\affiliation{Wako Nuclear Science Center, Institute of Particle and Nuclear Studies, High Energy Accelerator Research Organization, Wako 351-0198, Japan}

\date{\today}

\begin{abstract}
Nuclear mass is a key indicator of how the nuclear shell structure evolves. The recent mass measurement study of neutron-rich lanthanum isotopes [A. Jaries, $et~al$., Phys. Rev. Lett. {\bf 134}, 042501(2025)] reveals the presence of a distinct prominence in their two-neutron separation energies. However, its presence has been called into question based on the results of another mass determination [B. Liu, Ph.D. thesis, University of Notre Dame (2025)]. In this letter, we report an effort to clarify these contradictory results through the use of the simultaneous mass-lifetime measurement of the neutron-rich lanthanum isotope ${}^{149}$La using a multi-reflection time-of-flight mass spectrograph combined with a $\beta$-TOF detector. The peak corresponding to a $\beta$-decaying state was observed in the time-of-flight spectra at a position of $221(6)~{\rm keV/c^2}$ lighter than the reported ${}^{149}$La mass in A. Jaries, $et~al$., but our measured result is in excellent agreement with the mass value reported in B. Liu. We have concluded that this peak is the ground state of ${}^{149}$La. With this, the previously reported distinct prominence in the two-neutron separation energies disappears, while a new kink structure,   similar to that in the cerium isotopes, appears. Comparison with theoretical models suggests that a nuclear shape transition from octupole deformation to another type of deformation occurs around $N=91$ and is likely the cause of this kink structure.
\end{abstract}

\maketitle


Nuclear interactions can be sorted into two components: monopole and multipole parts.  The former corresponds to the effect of the spherical mean field, while the latter part majorly includes lower-order components, $i.e.$, quadrupole and octupole deformations. The competition between them determines the intrinsic nuclear shape. The transitions of nuclear shape can be found in sudden changes of the nuclear binding energy (BE), which serves as a key indicator for nuclear structure evolution.

The two-neutron separation energy, defined as $S_{\rm 2n} (N, Z) = {\rm BE}(N, Z) - {\rm BE}(N-2, Z)$, is one of the useful indicators to discuss nuclear structure evolution \citep{Lunney2003}. One of the important nuclear-physics quantities characterized by $S_{\rm 2n}$ is the neutron magic number, which appears as a drop in the $S_{2n}$ trend. 
In regions far from stability on the nuclear chart, traditional magic numbers vanish while new ones emerge. Direct mass measurements support the neutron magic number $N=32$ in calcium \citep{Wienholtz2013}, first suggested by $\beta$-end point measurements  \citep{Huck1985}. Further studies revealed magicities at $N=32$ and $N=34$ in the isotopes from potassium to scandium \citep{Rosenbusch2015, Michimasa2018, Meisel2020, Leistenschneider2021}. However, the charge radii measurements \citep{Ruiz2016, Koszorus2021} do not support the $N=32$ shell in potassium and calcium. For titanium and vanadium, a shell closure at $N=40$ was proposed \citep{Michimasa2020}, but it has recently been rejected \citep{Iimura2023}.

Distinct changes in $S_{2n}$ trends also indicate nuclear shape transitions and are observed at $N \approx 60$ and $N \approx 90$ on the neutron-rich side. The former was initially confirmed in zirconium with $N=58-60$ \citep{RintaAntila2004} and later across rubidium to molybdenum isotopes \citep{Hager2006, Rahaman2007, Hager2007, Simon2012, Klawitter2016, Lunney2025}. At $N \approx 90$, the bumps in the $S_{2n}$ trends for the rare-earth isotopes are observed by direct mass measurements \citep{Schelt2012, Orford2018, Orford2020, Orford2022, Kimura2024, Ray2024arXiv, Spataru2025}.

The rare-earth region with $N \approx 90$ is located near the area centered on the neutron-rich barium isotope ${}^{146}{\rm Ba}_{90}$, which is one of only two regions where the strongest octupole correlations are predicted (the other is around the neutron-deficient radium isotope ${}^{224}{\rm Ra}$) \citep{Butler2016}. This region is a suitable venue for examining nuclear theories due to the presence of various shape-related phenomena, such as shape coexistence and static octupole deformation. Thus, the experimental information on how nuclear deformation evolves around ${}^{146}{\rm Ba}$ is necessary to improve the understanding of the multipole contributions in the nuclear force.

\begin{figure}[t]
\includegraphics[width=0.475\textwidth,  bb = 0 0 362 725, clip, trim=0 0 0 0]{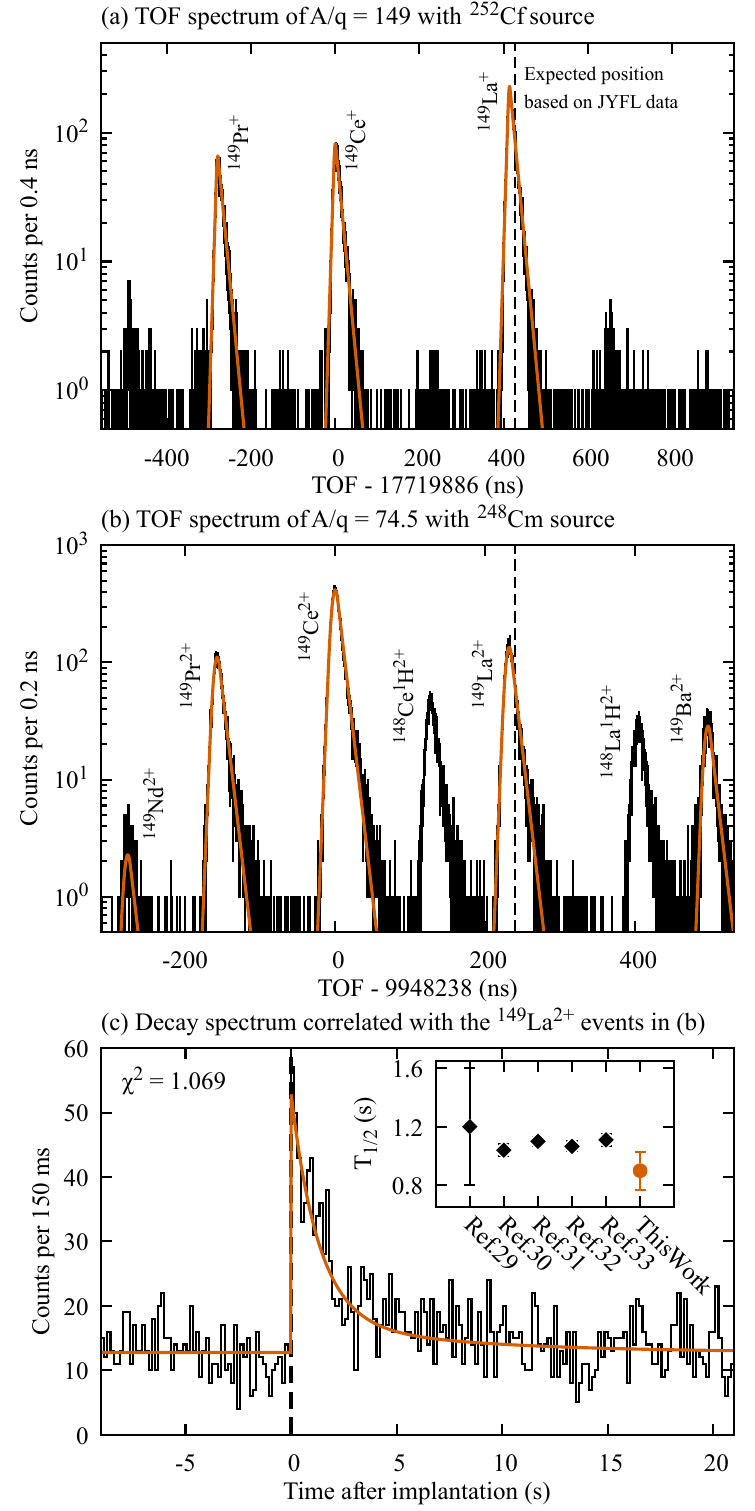}
\caption{(a) TOF spectrum of the singly-charged $A=149$ isobar series produced from the ${}^{252}{\rm Cf}$ source. The red-colored lines indicate the fit results. The vertical dashed line represents the expected peak position of ${}^{149}{\rm La}$ based on the JYFL result. (b) the same as (a), but for doubly-charged ions with the ${}^{248}{\rm Cm}$ source. (c) Decay spectrum of the TOF-event-correlated $\beta$-decay events in (b). A comparison of the ${}^{149}{\rm La}$ half-life with the literature values is also presented. For all panels, the red-colored lines indicate the fit results. \label{rslt149La}}
\end{figure}

\begin{figure}[t]
\includegraphics[width=0.475\textwidth,  bb = 0 0 362 181, clip, trim=0 0 0 0]{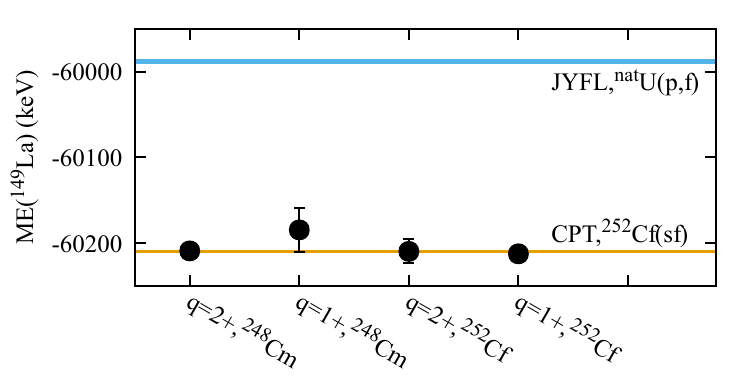}
\caption{The black dots show the measured mass excess values of ${}^{149}$La. Charge state and production method are labeled on the horizontal scale. The cyan- and orange-colored bands indicate the reported mass excess values by JYFL and CPT, respectively. \label{ME}}
\end{figure}

The recent advance in the region around ${}^{146}{\rm Ba}$ is the mass measurement of the neutron-rich lanthanum isotopes ${}^{148-153}{\rm La}_{91-96}$ with the JYFLTRAP double Penning trap spectrometer at the University of Jyv\"{a}skyl\"{a} \citep{Jaries2025} (hereafter denoted as ``JYFL''); this study reveals the presence of a prominent bump in the $S_{2n}$ plot of the neutron-rich lanthanum isotope series. The authors claimed that this prominence is unlike that in the cerium isotope series, and the evolution in $S_{2n}$ from $N=94$ to 95 is one of the strongest changes in the chart of nuclides in the region away from shell closures. However, another mass measurement study with the Canadian Penning Trap at CARIBU facility at Argonne National Laboratory \citep{Liuthesis} (``CPT") has reported a contradictory result for the atomic mass of ${}^{149}{\rm La}_{92}$; the CPT result is lighter by $221.6(32)~{\rm keV/c^2}$ from the mass reported in JYFL. Considering the high mass resolving power of Penning trap mass spectrometry, this difference is astounding. Since their measurements were carried out with different charge states, $q = 1+$ for JYFL and $q = 2+$ for CPT, it is possible that one group inadvertently has misidentified the ion. 
If the CPT result accurately reflects the mass of the ${}^{149}{\rm La}$ ground state, it significantly alters the $S_{\rm 2n}$ trend from that reported in JYFL and thus impacts the nuclear structure evolution in the neutron-rich lanthanum isotopes, which are situated in the mid-shell region.

We will point out the relation between this issue and the production methods. The discovery of ${}^{149}{\rm La}$ was reported by Engler $et~al.$ \citep{Engler1979} in 1979. Since then, several experimental half-lives have been reported; in all cases \citep{Engler1979, Warner1986, Reeder1986, Rudstam1993, Wu2017}, induced fission of uranium isotopes was used to produce it. Their results are quite similar, and their weighted average is the present adopted value \citep{ENSDF}. However, no half-life measurement with a spontaneous fission source had been reported. For its mass, as discussed above, two different values are reported by the JYFL and CPT groups. They employed different production methods: proton-induced fission of uranium ${}^{\rm nat}{\rm U} (p,{\rm f})$ for JYFL and spontaneous fission of californium ${}^{252}{\rm Cf}({\rm sf})$ for CPT. In addition, the prompt $\gamma$-ray measurement with ${}^{248}{\rm Cm}({\rm sf})$ pointed out that there eventually might be a long-lived isomeric state \citep{Urban2007}. Thus, it is possible that differences in production methods might account for the discrepancy between these mass determinations if a long-lived isomeric state exists.


The recent development of the $\alpha/\beta$-TOF (time-of-flight) detectors allows for simultaneous measurements of nuclear mass and decay properties \citep{Niwase2020, Niwase2023}. Our previous efforts in measuring the masses of superheavy \citep{Schury2021} and heavy nuclides \citep{Niwase2021}, combined with $\alpha$-decay properties, indicate that simultaneous measurements are useful for unambiguous event identification. In this letter, we demonstrate that the simultaneous mass-lifetime measurement solves the issue regarding the mass of the ${}^{149}{\rm La}$ ground state. 

The experiment was conducted using a multi-reflection time-of-flight mass spectrograph (MRTOF-MS) at RIKEN \citep{Schury2014} combined with a $\beta$-TOF detector \citep{Niwase2023}. The details of the experimental setup are available elsewhere \citep{Kimura2024}. The ${}^{149}{\rm La}$ ions were produced via spontaneous fission using two different sources: 9.25 MBq ${}^{252}{\rm Cf}$ and 67 kBq ${}^{248}{\rm Cm}$. The measurements were carried out for both singly and doubly charged ion states, for each fission source: the TOF spectra were taken with four different conditions in total.

The analysis strategy for extracting nuclear masses from TOFs is the same as used in \citep{Kimura2024}, and for the TOF correlated $\beta$-decay events, we refer to \citep{Kimura2024b}. Here, we reiterate the essential points of the analysis. The single reference method \citep{Ito2013} was employed for extracting mass values; the ion mass $m_{\rm X}$ is given by $m_{\rm X} = \rho^2 m_{\rm ref}$, where $m_{\rm ref}$ is the mass of the reference ion and $\rho$ is the TOF ratio defined by $\rho \equiv (t_{\rm X} -t_0) / (t_{\rm ref} -t_0)$. $t_{\rm X}$ and $t_{\rm ref}$ are the TOFs of the analyte and the mass reference, respectively, and $t_0$ is the systematic offset depending on the measurement system. The peaks were fitted with a modified exponential Gaussian hybrid function previously shown to provide high accuracy \citep{Kimura2024} to determine $\rho$. The isobaric mass references were used for mass determination. With this, the uncertainty caused by the error of the systematic offset $t_0$ is $ \delta (\rho^2)_{\rm sys} /\rho^2 \lesssim 10^{-9}$ \citep{Kimura2021} and is negligible in the present study. 

{
\tabcolsep = 0.5pt
\begin{table}[t]
\caption{Weighted-average mass excess values $\overline{\rm ME}$ \citep{SUPPL}. The ${}^{149}{\rm Ce}$ was used as the mass reference for all measurements. ME$_{\rm lit}$ indicates the literature's mass excess values, and $\Delta {\rm ME}$ represents the differences between ME$_{\rm lit}$ and the mass excess values of the present study: $\Delta \overline{\rm ME} \equiv \overline{\rm ME} - {\rm ME}_{\rm lit}$.  The detailed measurement results are given in the Supplemental Material. \label{WghtAveME}}
\renewcommand{\arraystretch}{1.5}
\begin{ruledtabular}
\begin{tabular}{cSScS}
\textrm{Nuclide}&
\multicolumn{1}{c}{\textrm{$\overline{\rm ME}$~(keV)}}&
\multicolumn{1}{c}{\textrm{ME$_{\rm lit.}$~(keV)}}&
\textrm{Ref. of lit.}&
\multicolumn{1}{c}{\textrm{$\Delta \overline{\rm ME}$}} 
\\ \hline
\colrule
${}^{149}{\rm Nd}$ &-74373(13)&-74375.5(21)& AME20 &3(14)\\
${}^{149}{\rm Pr}$ &-71059.5(52)&-71056.5(13)& CPT &-3(5)\\
${}^{149}{\rm La}$ &-60209.7(50)&-60209.9(16)& CPT &0(5)\\
 & &-59988.3(28)& JYFL &-221(6)\\
${}^{149}{\rm Ba}$ &-52821.5(8.4)&-52830.6(25)&AME20 &9(9)\\
\end{tabular}
\end{ruledtabular}
\renewcommand{\arraystretch}{1.0}
\end{table}
}

\begin{figure}[t]
\includegraphics[width=0.48\textwidth,  bb = 0 0 362 725, clip, trim=0 0 0 0]{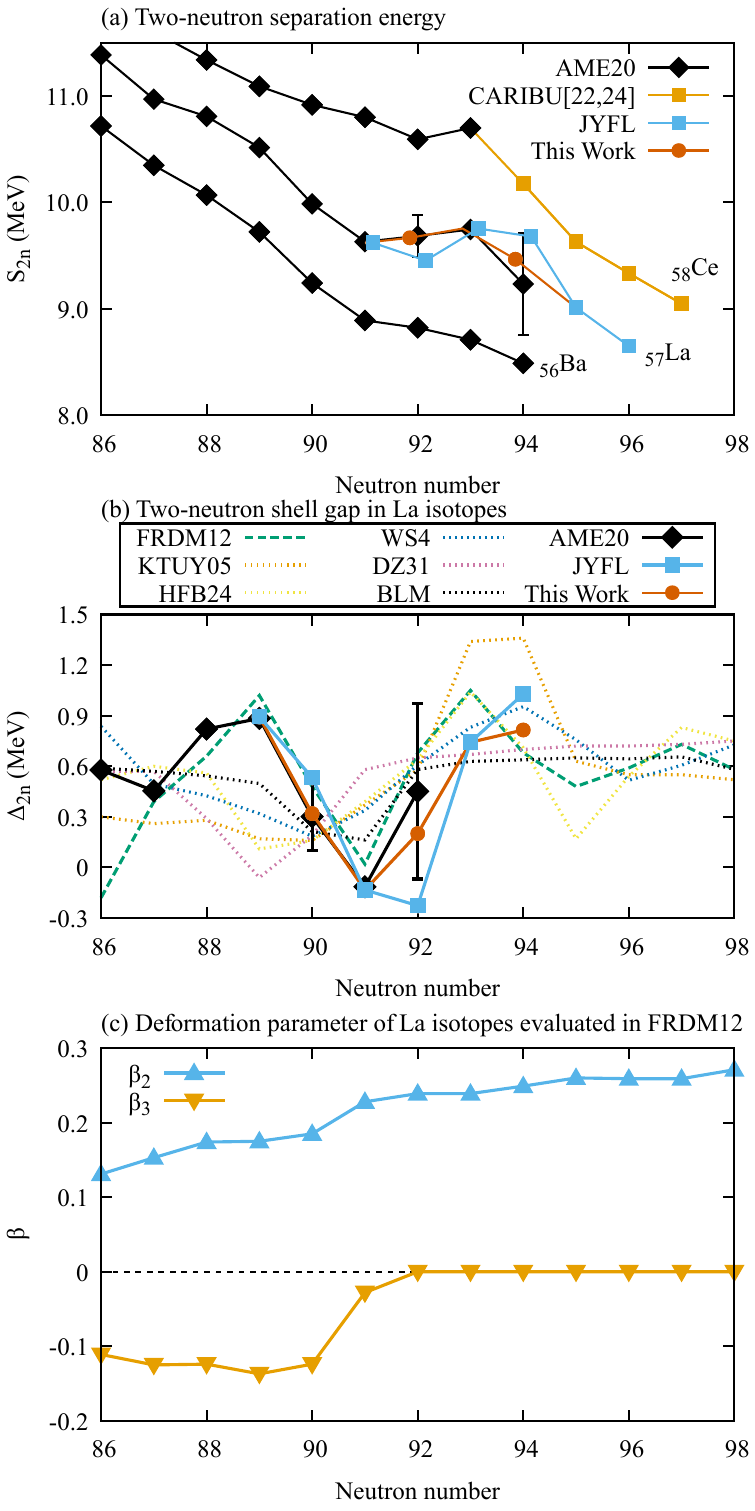}
\caption{(a) Two-neutron separation energies of the barium, lanthanum, and cerium isotope series. (b) Two-neutron shell gap of the lanthanum isotope series. For details regarding the theoretical predictions, see the main text. (c) The nuclear deformation parameters $\beta$ in the FRDM12. $\beta_2$ and $\beta_3$ correspond to the quadrupole and the octupole deformations, respectively. \label{S2nD2n}}
\end{figure}

The observed peak position of ${}^{149}{\rm La}$ is significantly different from the predicted position based on the JYFL result, as shown in Figs.~\ref{rslt149La} (a) and (b).  No visible sign of a second peak, which would correspond to the ions with the ${}^{149}{\rm La}$ mass measured by the JYFL group, was observed at the predicted position in any case. The analysis results of each measurement are presented in Fig.~\ref{ME}, and their weighted average values are given in Table~\ref{WghtAveME}. The updated mass value for ${}^{149}{\rm Ce}$ reported in CPT \citep{Liuthesis} was used as the mass reference. Taking all our measurements into account, the extracted mass excess (ME) value of ${}^{149}$La is ${\rm ME} = -60209.7(50)~{\rm keV}$, which is 221(6)~keV more bound than the JYFL result and in excellent agreement with the CPT result. For other nuclides present in our measurement, our results are consistent with the literature value within $1 \sigma$.

For another cross-check, the TOF-event-correlated $\beta$-decay events were analyzed for ${}^{149}{\rm La}{}^{2+}$ ions fed by the ${}^{248}{\rm Cm}$ source; the $\beta$-decay spectrum in Fig.~\ref{rslt149La}(c) shows that the observed peaks correspond to radioactive ions with a half-life of 0.90(13)~s, not molecules of stable isotopes. This value is in agreement with the literature value of 1.091(34) s \citep{ENSDF}, differing by less than $1.5\sigma$. The fit of the $\beta$-decay spectrum comprises a decay curve, which includes the contribution from the daughter nuclide, and a constant background. The half-life of daughter nuclide ${}^{149}{\rm Ce}$ was the fixed parameter in the fit process, and its value was taken from \citep{ENSDF}. The relative detection efficiency of ${}^{149}{\rm Ce}$ was estimated through simulation using Geant4 \cite{Agostinelli2003}. As a result, it ranges from 0.53 to 0.62, and the uncertainty of the extracted half-life includes the influence of this ambiguity. The total count rate of the $A=149$ isobar series at the $\beta$-TOF detector was around 0.06~cps during the measurement.

Based on these measurement results, we can conclude that the observed ${}^{149}{\rm La}$ corresponds to a ground state. In contrast, the present measurement results suggest that the state previously studied in JYFL might correspond to a long-lived isomeric state, assuming that spontaneous and induced fissions can produce different final states of the same fragments.

The new mass value of ${}^{149}{\rm La}$ significantly alters the $S_{2n}$ trend of the lanthanum isotopes at $N=92$ and 94 reported by JYFL, as shown in Fig.~\ref{S2nD2n} (a). The previously reported distinct prominence around $N=93$ has disappeared, while a kink structure appears at the same point. Then, the $S_{2n}$ trend of lanthanum isotopes now shows a similar pattern to that of cerium isotopes. 

The kink structure in the $S_{2n}$ plot of lanthanum isotopes can be understood as a result of a nuclear shape transition from octupole to another type of deformation. The two-neutron shell gap of the lanthanum isotopes, defined by $\Delta_{\rm 2n} (N, Z)= S_{\rm 2n}(N, Z) - S_{\rm 2n}(N+2, Z)$  \citep{Lunney2003}, is plotted in Fig.~\ref{S2nD2n}~(b) alongside several theoretical predictions: the finite-range liquid-drop model (FRDM12) \citep{Moller2016}, the model of Koura-Tachibana-Uno-Yamada (KTUY05) \citep{Koura2005}, the Skyrme-Hartree-Fock-Bogoliubov mass formula with the BSk24 parameter set (HFB24) \citep{Goriely2013}, the macroscopic-microscopic model with surface diffuseness correction (WS4) \citep{Wang2014}, the Duflo-Zuker parametrized mass formula with 31 parameters (DZ31) \citep{Duflo1995}, and the Bayesian machine learning mass model (BML) \citep{Niu2022}. The valley-like shape can be confirmed in the $\Delta_{\rm 2n}$ plot, and the bottom reaches negative values. This indicates a significant change in the nuclear shell structure, observed at $N=91$ from the CPT and the present measurements, whereas the result from JYFL shows them at $N=91$ and 92. Among the theoretical models, the FRDM12 well reproduces this trend in $\Delta_{\rm 2n}$ in the range of $87 \leq N \leq 94$. Figure~\ref{S2nD2n}~(c) shows the nuclear deformation parameters calculated in the FRDM12. One notable aspect in the evolution of deformation parameters is the disappearance of the octupole component $\beta_3$ around $N=91$. In contrast, the quadrupole component $\beta_2$ slightly enhances at the same point. The prompt $\gamma$-ray measurements of ${}^{145,147}{\rm La}$ indicate that octupole correlations weaken as the neutron number increases from $N=90$ to 92 \citep{Urban1996}, which is consistent with the trend of deformation parameters in the FRDM12. Neither level schemes nor spectroscopic information are available for the lanthanum isotopes of $N \ge 93$. For ${}^{148,150}{\rm Ba}_{92,94}$, the cores of ${}^{149,151}{\rm La}$, the presence of an increasing collectivity towards prolate deformation is pointed out, but the definite octupole character still remains \citep{Lica2018}. With this situation, their type of deformation has been unclear.  Therefore, one can conclude that a nuclear shape transition from octupole deformation to another type of deformation is the main cause of the behaviors of $S_{2n}$ and $\Delta_{\rm 2n}$ around $N=91$.

By considering the presence of the two $\beta$-decaying states, the issue with the spin-parity assignment in the level scheme of ${}^{149}{\rm La}$ could be resolved. It has been studied using two different methods: the $\beta$-delayed $\gamma$-ray spectroscopy of ${}^{149}{\rm Ba}$ \citep{Syntfeld2004} (``Sy04") and the prompt $\gamma$-ray measurements of ${}^{248}{\rm Cm}({\rm sf})$ \citep{Urban2007} (``Ur07"). The observed $\gamma$-transitions belong to different bands, and no overlap is confirmed. In Ur07, the authors conclude that the $\gamma$-transition by ${}^{248}{\rm Cm}({\rm sf})$ ends at the $7/2^-$ state, based on the nuclear level evolution of the odd-$A$ lanthanum isotopes. However, the other results from Sy04 and the measured ${}^{149}{\rm Ce}$ level scheme following the $\beta$-decay of ${}^{149}{\rm La}$ \citep{Syntfeld2002} (which employed ${}^{235}{\rm U} (n,{\rm f})$ for production) suggest the ground state spin of $3/2$. With this, Ur07 also points out the possibility that this $7/2^-$ state could be a long-lived isomeric state with an excitation energy of less than 35~keV; however, such a low-lying isomeric state has not yet been observed. In the present measurement, we use ${}^{248}{\rm Cm}({\rm sf})$ as the production method; therefore, the newly identified $\beta$-decaying state should be the same as the $7/2^-$ state observed in Ur07. Then the $7/2^-$ state is located lower than the $3/2$ state, and the constraint on the excitation energy of the $7/2^-$ state is now unnecessary. Consequently, a tentative spin assignment, which is consistent with all previous experimental results concerning the ${}^{149}{\rm La}$ level scheme, becomes available.


In summary, we conducted a simultaneous mass-lifetime measurement of ${}^{149}{\rm La}$. The peak corresponding to a $\beta$-decaying state was observed in the TOF spectra at a position of $221(6)~{\rm keV/c^2}$ lighter than the recent JYFL result. Therefore, we can conclude that the observed ${}^{149}{\rm La}$ corresponds to its ground state. With this, the distinct prominence in the $S_{\rm 2n}$ plot in the lanthanum isotope series disappears, replaced by a kink structure around $N=92$, which is similar to that in the cerium isotopes. The FRDM12 reproduces the trend of lanthanum isotopes' $\Delta_{\rm 2n}$. This suggests that a nuclear shape transition from octupole deformation to another type of deformation occurs around $N=91$ and is likely the cause of this kink structure. \\

\textit{Acknowledgments}--This work was supported by the Japan Society for the Promotion of Science, KAKENHI, Grants No. 17H06090, No. 22H04946, No. 23K13132, and No. 23K13137.

\bibliography{new_gs_149La}

\section{SUPPLEMENTAL MATERIALS}
The measured square of the TOF ratio $\rho^2$ of each measurement is tabulated in Table~\ref{Summary}. 

\begin{table*}[h]
\caption{Measured square of time-of-flight ratio $\rho^2$ and the extracted mass excess values ME. The ions of ${}^{149}{\rm Ce}^{n+}~(n=1,2)$ were used as the refernce of $\rho$. ${\rm ME}({}^{149}{\rm Ce}) = -66674.0(12)~{\rm keV}$ \citep{Liuthesis} was adopted for calculating the ME values listed below. ME$_{\rm lit}$ indicates the literature's mass excess values, and $\Delta {\rm ME}$ represents the differences between ME$_{\rm lit}$ and the mass excess values of the present study: $\Delta {\rm ME} \equiv {\rm ME} - {\rm ME}_{\rm lit}$.  \label{Summary}}
\renewcommand{\arraystretch}{1.5}
\begin{ruledtabular}
\begin{tabular}{clcSScS}
\textrm{Species}&
\multicolumn{1}{c}{\textrm{$\rho^2$}}&
\multicolumn{1}{c}{\textrm{$\delta (\rho^2) / \rho^2$}}&
\multicolumn{1}{c}{\textrm{ME~(keV)}}&
\multicolumn{1}{c}{\textrm{ME$_{\rm lit.}$~(keV)}}&
\textrm{Ref. of lit.}&
\multicolumn{1}{c}{\textrm{$\Delta {\rm ME}$~(keV)}} 
\\ \hline
\colrule
\multicolumn{2}{l}{${}^{248}{\rm Cm}$ source, $q = 2+$} & & & & & \\
${}^{149}{\rm Nd}^{2+}$ & 0.99994442(20)&$2.0 \times 10^{-7}$&-74384(28)&-74375.5(21)& AME20 &-8(28)\\
${}^{149}{\rm Pr}^{2+}$ & 0.999968367(56)&$5.6 \times 10^{-8}$&-71062.3(79)&-71056.5(13)& CPT &-6(8)\\
${}^{149}{\rm La}^{2+}$ & 1.000046602(56)&$5.6 \times 10^{-8}$&-60209.2(78)&-60209.9(16)& CPT &1(8)\\
${}^{149}{\rm Ba}^{2+}$ & 1.000099854(62)&$6.2 \times 10^{-8}$&-52821.7(86)&-52830.6(25)&AME20 &-9(9)\\
& & & & & & \\
\multicolumn{2}{l}{${}^{248}{\rm Cm}$ source, $q = 1+$} & & & & & \\
${}^{149}{\rm Pr}^{+}$ & 0.99996840(19)&$1.9 \times 10^{-7}$&-71058(26)&-71056.5(13)& CPT &-1(26)\\
${}^{149}{\rm La}^{+}$ & 1.00004678(19)&$1.9 \times 10^{-7}$&-60185(26)&-60209.9(16)& CPT &25(26)\\
& & & & & & \\
\multicolumn{2}{l}{${}^{252}{\rm Cf}$ source, $q = 2+$} & & & & & \\
${}^{149}{\rm Nd}^{2+}$ & 0.99994453(11)&$1.1 \times 10^{-7}$&-74370(15)&-74375.5(21)& AME20 &6(15)\\
${}^{149}{\rm Pr}^{2+}$ & 0.999968505(98)&$9.8 \times 10^{-8}$&-71043(14)&-71056.5(13)& CPT &13(14)\\
${}^{149}{\rm La}^{2+}$ & 1.00004660(10)&$1.0 \times 10^{-7}$&-60210(14)&-60209.9(16)& CPT &0(14)\\
${}^{149}{\rm Ba}^{2+}$ & 1.00009990(30)&$3.0 \times 10^{-7}$&-52815(42)&-52830.6(25)&AME20 &15(42)\\
& & & & & & \\
\multicolumn{2}{l}{${}^{252}{\rm Cf}$ source, $q = 1+$} & & & & & \\
${}^{149}{\rm Pr}^{+}$ & 0.999968374(58)&$5.8 \times 10^{-8}$&-71062.2(81)&-71056.5(13)& CPT &-6(8)\\
${}^{149}{\rm La}^{+}$ & 1.000046577(55)&$5.6 \times 10^{-8}$&-60212.6(78)&-60209.9(16)& CPT &-3(8)\\
\end{tabular}
\end{ruledtabular}
\renewcommand{\arraystretch}{1.0}
\end{table*}

\end{document}